\documentclass{osa-article}

\journal{osajournal}

\articletype{Research Article}

\usepackage{graphicx}
\usepackage{dcolumn}
\usepackage{bm}
\usepackage{xcolor}

\begin{document}

\title{Large-scale photonic natural language processing}

\author{Carlo Michele Valensise\authormark{1}, Ivana Grecco\authormark{2}, Davide Pierangeli\authormark{3,2,1,*}, and Claudio Conti\authormark{2,3,1}}

\address{
\authormark{1} Enrico Fermi Research Center (CREF), 00184 Rome, Italy\\
\authormark{2} Physics Department, Sapienza University of Rome, 00185 Rome, Italy\\
\authormark{3} Institute for Complex Systems, National Research Council (ISC-CNR), 00185 Rome, Italy\\
}
\email{\authormark{*}davide.pierangeli@roma1.infn.it}

\begin{abstract}
Modern machine learning applications require huge artificial networks demanding in computational power and memory.
Light-based platforms promise ultra-fast and energy-efficient hardware, which may help in realizing next-generation data processing devices.
However, current photonic networks are limited by the number of input-output nodes that can be processed in a single shot.
This restricted network capacity prevents their application to relevant large-scale problems such as natural language processing.
Here, we realize a photonic processor with a capacity exceeding $1.5 \times 10^{10}$ optical nodes,
more than one order of magnitude larger than any previous implementation, 
which enables photonic large-scale text encoding and classification.
By exploiting the full three-dimensional structure of the optical field propagating in free space,
we overcome the interpolation threshold and reach the over-parametrized region of machine learning,
a condition that allows high-performance natural language processing with a minimal fraction of training points. 
Our results provide a novel solution to scale-up light-driven computing and open the route to photonic language processing.
\end{abstract}

\section{INTRODUCTION}

Advanced artificial intelligence systems are becoming extremely demanding in terms of training time
and energy consumption \cite{Narayanan2021, Strubell2020}, since an ever-larger number of trainable parameters is required
to exploit over-parametrization and achieve state-of-the-art performances \cite{Chatelain2021, Thompson2020}.
In turn, large-scale, energy-efficient computational hardware is becoming a subject of intense interest. 

Photonic neuromorphic computing systems \cite{Wetzstein2020, Shastri2021, Zhou2022} offer large throughput \cite{Xu2021} and energetically efficient \cite{McMahon2021} hardware accelerators, based on integrated silicon photonic circuits \cite{Shen2017, Tait2017, Feldmann2019, Stelzer2021, Feldmann2021}, engineered meta-materials \cite{Engheta2019}, or 3D printed diffractive masks \cite{Lin2018, Zhou2021}. 
Other light-based computing architectures leverage Reservoir Computing (RC)~\cite{Verstraeten2007}
and Extreme Learning Machine (ELM)~\cite{Huang2012} computational paradigms, 
in which the input data is mapped into a feature space through a fixed set of random weights and training is performed only on the readout linear layer. 
Optical reservoir computers \cite{VanDerSande2017, Brunner2013, Vinkier2015, Larger2017, Ballarini2020, Saade2016, Bueno2018, Antonik2019, Rohm2020, Paudel2020, Miscuglio2020, Sunada2021, Pavesi2021, Porte2021} and Photonic Extreme Learning Machines (PELMs) \cite{Pierangeli2021, Psaltis2021, Pierangeli2021_2, Ciuti2022} applies successfully to various learning tasks, 
ranging from time series prediction \cite{Rafayelyan2020, Ortin2015} to image classification \cite{Matuszewski2021,  Lupo2021}.
In spite of the remarkable performances achieved by these architectures, 
their impact on large-scale problems is limited, mainly due to size constraints.

Here, we demonstrate photonic machine learning at an ultra-large scale
by mapping the optical output layer in the full three-dimensional (3D) structure of the optical field.
This original approach offers inherent scalability to the optical device and 
effortless access to the over-parametrized learning region. 
The 3D-PELM that we implement is able to process simultaneously up to 250000 total input-output nodes via spatial light modulation,
featuring a total network capacity one order of magnitude larger than existing optical reservoir computers. 
Our large-scale photonic network allows us to optically implement a massive text classification problem,
the sentiment analysis of the Image Movie Database (IMDb) \cite{Maas2011}, and enable, for the first time to our knowledge,
the observation of the double descent phenomenon \cite{Belkin2019, Advani2020, Montanari2020} on a photonic computing device.
We demonstrate that, thanks to its huge number of optical output nodes, 
the 3D-PELM successfully classifies text by using only a limited number of training points,
realizing energy-efficient large-scale text processing.

\begin{figure*}[t!]
\centering
\vspace*{-0.1cm}
\hspace*{-1.6cm} 
\includegraphics[width=1.25\columnwidth]{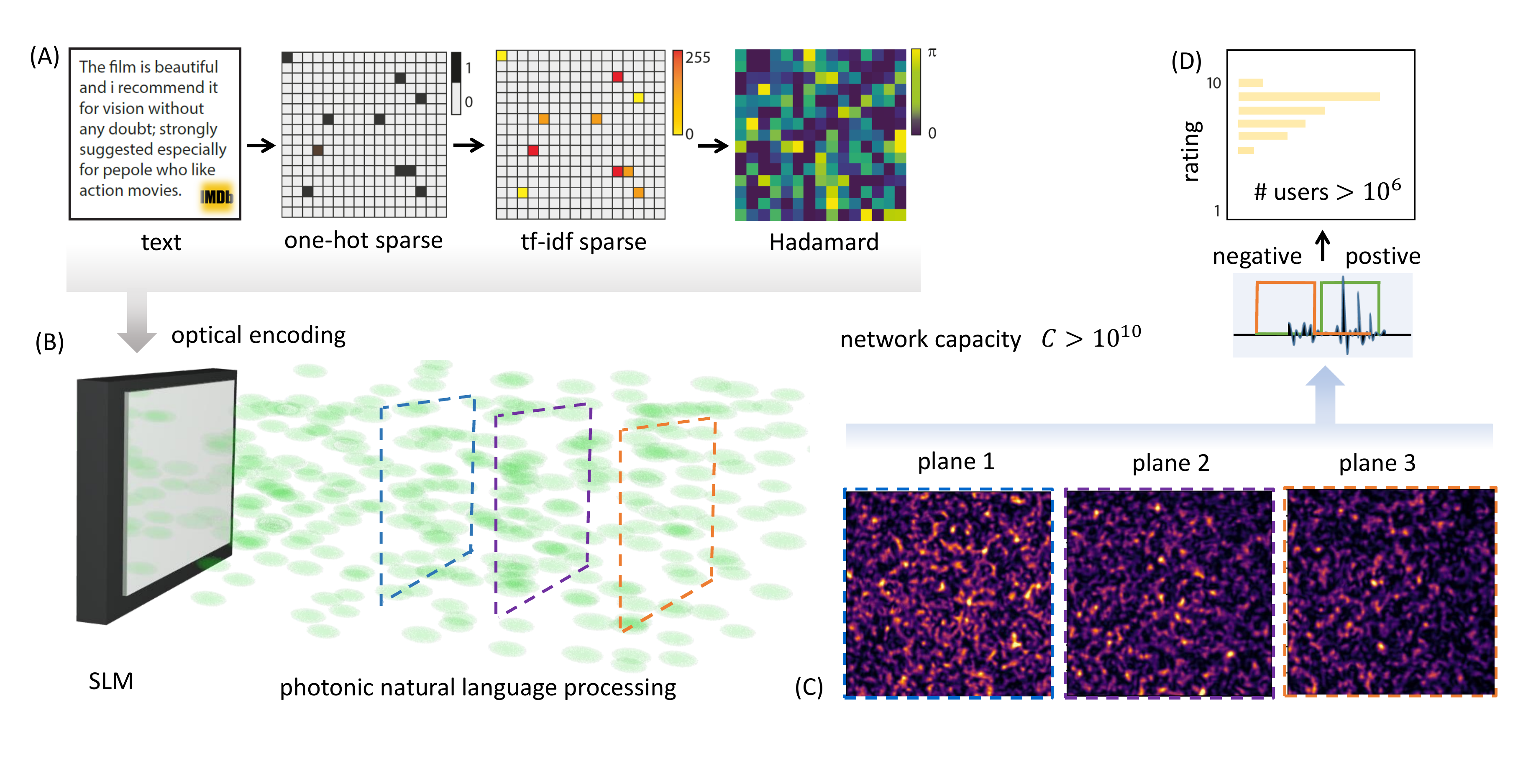} 
\vspace*{-0.8cm}
\caption{ {\bf Three-dimensional PELM for natural language processing.}
 (A) The text database entry is a paragraph of variable length. Text pre-processing: a sparse representation of the input paragraph 
is mapped into an Hadamard matrix with phase values in $[0,\pi]$. (B) The mask is encoded into the optical wavefront by a phase-only spatial light modulator (SLM). Free-space propagation of the optical field maps the input data into a 3D intensity distribution (speckle-like volume).
(C) Sampling the propagating laser beam in multiple far-field planes enables the up-scaling of the corresponding feature space.
(D) Intensities picked from all the spatial modes form the output layer $\mathbf{H_{\mathrm{3D}}}$ that undergoes training via ridge regression. 
By using three planes ($j=3)$ we get a network capacity $C>10^{10}$.
The inset example shows a binary text classification problem for large-scale rating.
} 
\vspace{-0.1cm}
\label{Figure1}
\end{figure*}

\section{RESULTS}

\subsection*{ \textbf{Three-dimensional PELM}}

A PELM classifies a dataset with $N$ points $\mathbf{X}=\left[ X_1,...,X_N \right]$
by mapping the input sample $X_i  \in \mathbb{R}^L$ into a high-dimensional feature space through a nonlinear transformation 
that is performed by a photonic platform \cite{Pierangeli2021}.
To perform the classification, the output matrix $\mathbf{H}$, which contains the measured optical signals, 
is linearly combined with a set of $M$ trainable readout weights $\mathbf {\beta}$ (see Appendix~A).
In all previous PELM realizations \cite{Saade2016, Pierangeli2021, Psaltis2021, Ortin2015, Lupo2021}, 
the intensities stored in $\mathbf{H}$ only contain information on the optical field at a single spatial or temporal plane. 
Here, to scale up the optical network, we exploit the entire three-dimensional optical field.
Figure 1(B) shows a schematic of our 3D-PELM: the photonic network uses as output nodes the full 3D structure of the optical field
propagating in free space. The transformation of the input data sample into an optical intensity volume
occurs through linear wave mixing by coherent free-space propagation and intensity detection. Specifically,
the 3D optical mapping reads as 
\begin{equation}
\mathbf{H_{\mathrm{3D}}}=\sum_{j} \mathbf{H}_j ,
\end{equation}
\begin{equation}
\mathbf{H}_j = G_{j}(\mathbf{M}_j\,\exp{i(\mathbf{X}+\mathbf{W})})\,,
\end{equation}
where $\mathbf M_j$ is the transfer matrix that models coherent light propagation to the $j$th detection plane, e.g.,
the discretized diffraction operator.
The $\mathbf M_j$ have complex coefficients, and thus the photonic scheme implements a complex-valued neural network \cite{Marcucci2020, Zhang2021}.
In Eq.~(2) the input dataset is encoded by phase modulation, while $\mathbf{W}$ is a pre-selected random matrix that embeds the input signal and provides
network biases. $G_{j}$ is the response function of the $j$th detector to the impinging light field.
In experiments, to collect uncorrelated output data $\mathbf{H}_{j}$, we employ multiple cameras that simultaneously image 
the intensity in distinct far-field planes (see Appendix~B).
These intensity distributions are uncorrelated in order to acquire non-redundant information.
Therefore, the speckle-like pattern detected by each camera (Fig. 1(C)) represents 
the optical projection of the $X_i$ input data on a distinct portion of the entire feature space.

\subsection*{\textbf{Optical encoding of natural language sentences}}

The NLP task we considered is the classification of the IMDB dataset, constructed by {\it Maas et al.} \cite{Maas2011}.
To encode natural language on the optical setup, we need to develop an optics-specific representation of the text data.  
Since text classification tasks have remained elusive in photonic neuromorphic computing,
we here introduce the problem. 
A similar issue is crucial also in digital NLP, where the text encoding problem consists in obtaining a numerical 
representation of the input that is convenient for digital learning architectures \cite{Babic2020}.
Specifically, given a vocabulary, i.e., the set of all unique words in the corpus, a basic encoding method is the one-hot technique,
in which each word corresponds to a boolean vector with vocabulary size $V$.
Consequently, paragraph of different lengths can be represented by boolean vectors of size $V$,
with the element $x_{ki}=1$ that indicates the presence of the $k$th vocabulary word in the paragraph $X_i$  [Fig. 1(A)].
An input dataset with $N$ samples, in which each database entry is a text paragraph [see Fig. 1(A)], thus becomes a $N \times V$ matrix.   
The dimension of the single entry $X_i$ to encode thus scales with the number of words $V$ in the vocabulary ($10^4-10^5$). 
Photonic text processing hence necessitates an optical platform able of encoding huge input data.
Given the large number of input modes $L$ supported by spatial light modulation, our 3D-PELM is the most convenient scheme for the scope.  

Optical encoding of natural language in the 3D-PELM requires using a SLM with a fixed number of input modes for sentences with variable lengths.
We can thus follow the one-hot method employed in digital NLP.
We leverage an extension of one-hot encoding, the so-called tf-idf representation, 
and construct the input representation $\mathbf{X}_{\text{tfidf}}$. In information retrieval,
the $\operatorname{tf-idf}(i,j)$ statistics reflect how important is a word $k$ to a document $j$ (frequency $n_{kj}$) in a collection of documents.
It is defined as: $\operatorname{tf-idf}(k,j)=\frac{n_{kj}}{V}\cdot \log_{10} \left( \frac{N}{S_k}\right)\,$,
where $V$ is the length of a paragraph, and $S_k$ is the number of sentences containing the $k$th word. 
Within the tf-idf representation, each paragraph becomes a sparse, real vector [see Fig. 1(A)]
whose non-zero elements are the tf-idf values of the words composing it. 
To optically encode these sparse data, 
we applied to $\mathbf{X}_{\text{tfidf}}$ the Walsh-Hadamard transform (WHT) \cite{Ashrafi2017}, 
which decomposes a discrete function in a superposition of Walsh functions. 
The transformed dataset $\mathbf{X}_{\text{WHT}}$ 
is a dense matrix [Fig. 1(A)], which is displayed on the SLM as a phase mask within the [0,$\pi$] range
($\mathbf{X}=\mathbf{X}_{\text{WHT}}$ in Eq. (2)).  
By exploiting the large number of available input pixels, we encode paragraphs with a massive size $V$
that contain hundreds of thousands of words.

\subsection*{\textbf{Observation of the photonic double descent}}

\begin{figure}[t!]
\centering
\hspace*{-0.8cm} 
\includegraphics[]{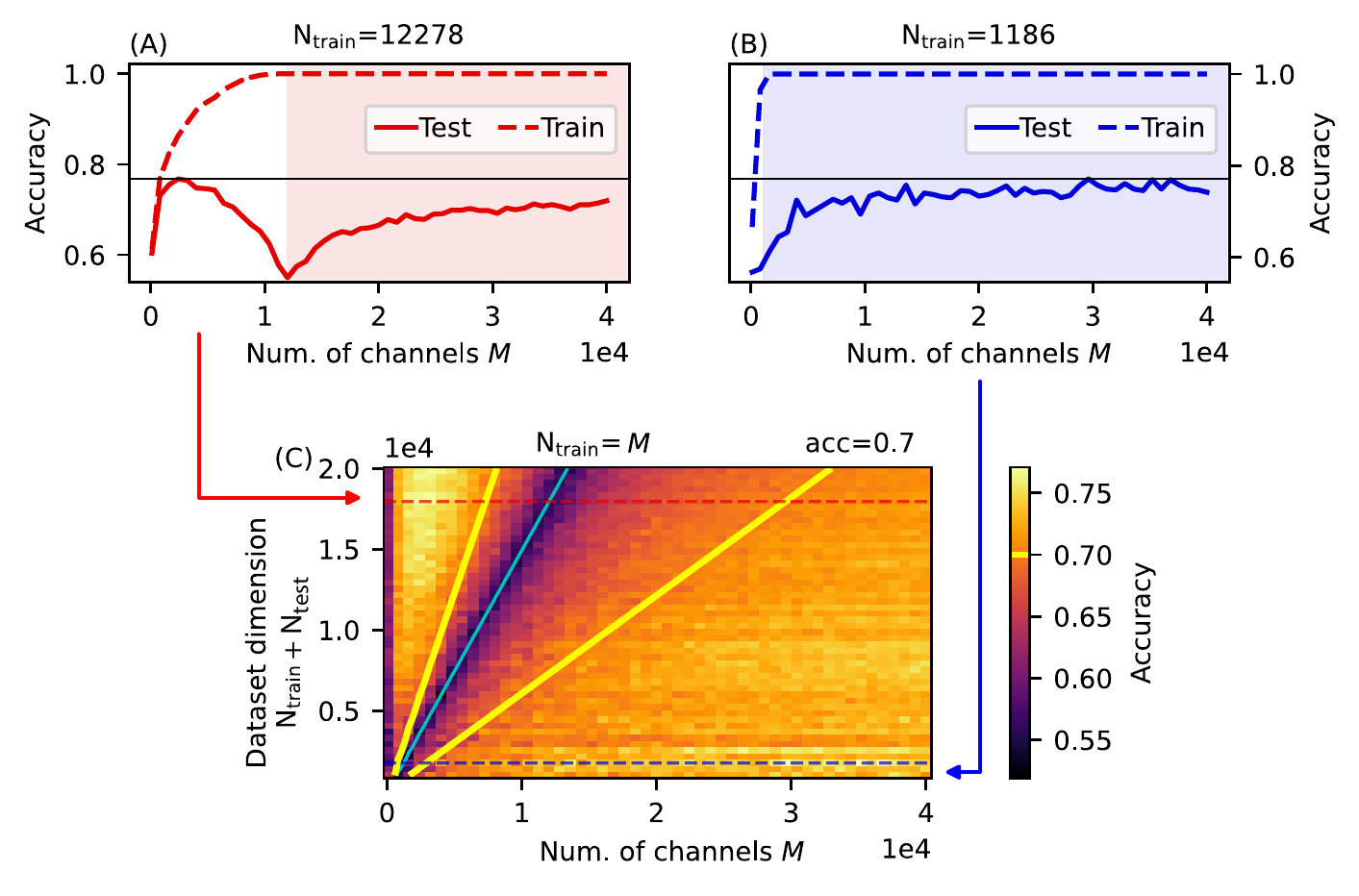}
\caption{Photonic NLP.
(A-B) Train and test accuracy of the 3D-PELM on the IMDb dataset as a function of the number of output channels. 
The shaded area corresponds to the over-parametrized region. 
The configuration in (B) allows reaching very high accuracy in the over-parametrized region with a dataset limited to $N_{\text{train}}= 1186$ training points. In (A) the same accuracy is reached in the under-parametrized region with $N_{\text{train}}= 12278$.
Black horizontal lines correspond to the maximum test accuracy achieved ($0.77$). 
(C) IMDb classification accuracy by varying the number of features $M$ and training dataset size $N_{\text{train}}$. 
The boundary between the under and over-parametrized region (interpolation threshold),
 $N_{\text{train}}=M$, is characterized by a sharp accuracy drop (cyan contour line).}
\label{Figure2}
\end{figure}

Machine learning models with a very large number of trainable parameters show unique features with respect to smaller models, 
such as the double descent phenomenon. This effect is a resonance in the neural network performance, ruled by the ratio between the number of trainable parameters $M$ and the number of training points $N_{\text{train}}$. 
In the under-parametrized regime ($M<N_{\text{train}}$), good performances are obtained by balancing model bias and variance.
As $M$ grows, models tend to overfit training data, and prediction performances get worse until the interpolation threshold ($M=N_{\text{train}}$) is reached, a point where the model optimally interpolates the training data and the prediction error is maximum. 
Beyond this resonance, in the over-parametrized regime, the model keeps interpolating training points, 
but performances on the test set reach the global optimum \cite{Soltanolkotabi2019}. 

We experimentally implement photonic NLP and investigate the double descent effect on our large-scale 3D-PELM. 
Specifically, we analyze the classification accuracy for the IMDb task (Appendix~D) 
as a function of the features $M$ and training set size $N_{\text{train}}$.
In Figure~2(A) we report the observation of the double descent. 
We observe a dip in test accuracy as the number of channels $M$ reaches the number of training points $N_{\text{train}}$. 
Beyond this resonance, also known as interpolation threshold, 
we find the over-parametrized region in which maximum accuracy on the training set is achieved.
The behavior is obtained via training on $N_{\text{train}}\simeq 1.2\times10^4$ examples and using
fixed train/test split ratio of $0.67/0.33$. In this case, the larger classification accuracy
is found in the under-parametrized region (0.77).
Conversely, in Figure~2(B), we consider a much smaller number of training points, $N_{\text{train}}\simeq 10^3$.
Remarkably, we reach the same optimal accuracy despite we are using only a fraction of the available training points. 
This observation reveals a particularly favourable learning region of the 3D-PELM 
in which only a reduced number of training points is required to reach high-accuracy levels.
In fact, from the operational standpoint, 
it is much more efficient to measure in parallel a large number of modes rather than sequentially processing many training examples. 
In Figure~2(C) we report the full dynamics of the double descent phenomenon, by continuously varying both $M$ and $N_{\text{train}}$.
We observe the accuracy dip shifts with a constant velocity.
This result shows the existence of an optimal learning region that is accessible on the 3D-PELM
thanks to its large number of photonic nodes.

\subsection*{\textbf{NLP at ultra-large scale}}

\begin{figure}[t!]
\centering
\hspace*{-0.4cm}
\includegraphics[]{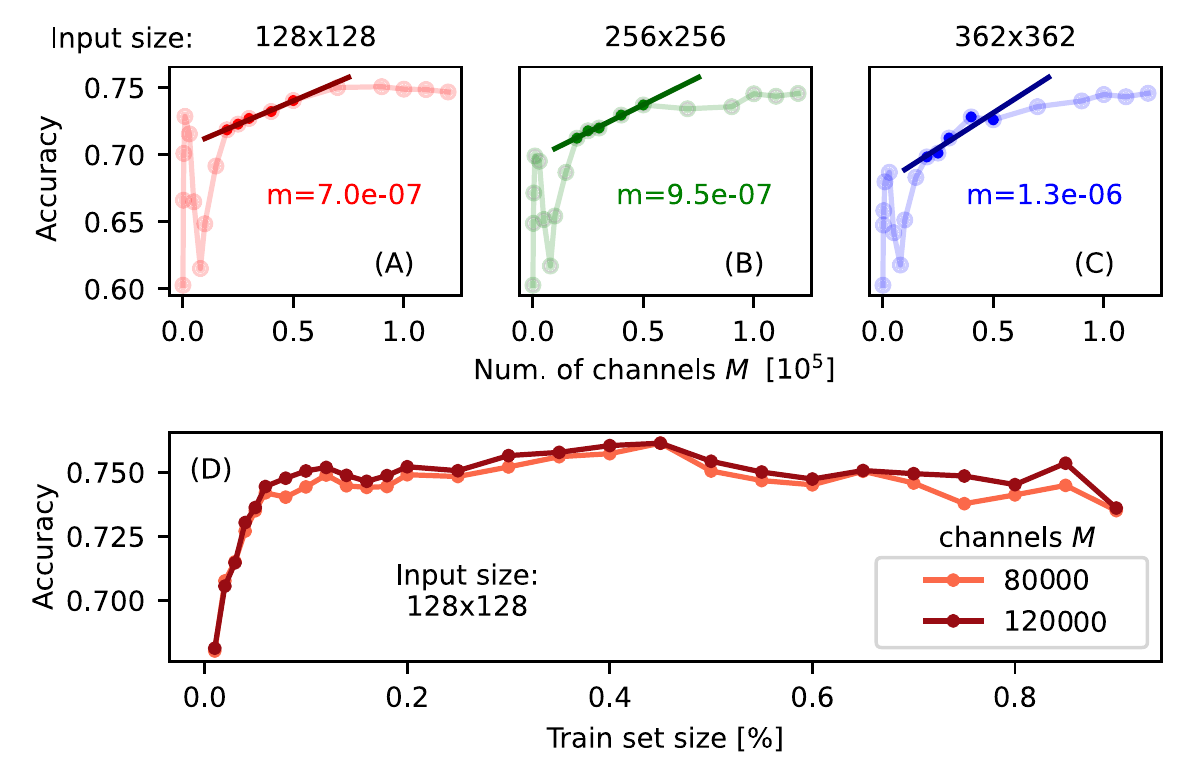}
\caption{Performances at ultra-large scale. (A-C) Accuracy as a function of $M$ for different input sizes $L$.
In all cases, the 3D-PELM performance saturates in the over-parameterized region reaching a plateau.  
A linear fit of the data preceding the plateau shows that the onset of the saturation is faster for datasets with a larger input space.
The corresponding angular coefficient $m$ is inset in each panel. 
(D) Test accuracy varying the training set size, for $M=0.8\times10^5$ and $M=1.2\times10^5$.}
\label{Figure3}
\end{figure}

The observed operational advantage of the over-parametrized region indicates that by further increasing the number of output modes
one can increase training effectiveness and/or performances \cite{Chatelain2021}.
In our 3D-PELM we reached $M=120000$ readout channels that are independent from each other. 
We also considered larger input spaces, extending the vocabulary so as to include the whole set of words in the corpus ($V\simeq 2^{17}$).
Figure~3(A) shows that the test accuracy reaches a plateau as $M$ increases in the over-parametrized region.
Saturation indicates that all the essential information encoded within the optical field has been extracted through the available channels.
However, we observe a change in the performance as we employ more input features $L$, Fig. 3(B-C).
To estimate the rate by which performance improves as more channels were used for training,
we estimate the angular coefficient $m$ of the linear growth that precedes the plateau. 
Although the onset of saturation can be estimated using diverse criteria, 
and $m$ varies depending on the measured range, its trend when increasing $L$ remain unaltered.
We note a relevant enhancement of $m$ for $L=362\times 362$,
which indicates that PELMs featuring larger input spaces are able to reach optimal performances with a lower number of parameters.
Figure~3(D) reports the accuracy as a function of the training-test split ratio (keeping fixed $N_{\text{train}}=6.7\times10^3$) 
for $M=0.8\times10^5$ and $M=1.2\times10^5$.
Importantly, a limited number of examples ($\sim20\%$) is enough to reach maximum accuracy in the over-parametrized region.

\subsection*{\textbf{Optical network capacity}}

To establish a comparison among the various photonic neuromorphic computing devices,
we introduce the optical network capacity as a useful ingredient related to the over-parametrization context.
We define the capacity $\mathcal C$ of a generic optical neural network as the product between the number of input and output nodes
that can be processed in a single iteration, $\mathcal C=L\times M$.
This quantity gives direct information on the kind of large-scale problems that can be implemented on the optical setup.
It depends only the number of controllable nodes, which is the quantity that is believed to play the main role in big data processing \cite{Chatelain2021}. Specifically, $L$ sets the size of the dataset that can be encoded, 
while $M$ reflects the embedding capacity of the network, as larger dataset necessitate larger feature spaces to be learned.
Moreover, $\mathcal C$ also furnishes an indication on how far is the over-parametrized regime for a given task.
Useful over-parametrized conditions can be reached only if $C\gg N\times L$. 
We remark that the capacity is not a measure of the processor accuracy.
It is instead a useful quantity to compare the scalability of different photonic computing devices.

In Tab.~1 we report the optical network capacity for various photonic processors that have been recently demonstrated.
We focus on photonic platforms that exploit RC and ELM paradigms, since these devices are suitable for big data processing  
and may present the so-called photonic advantage at a large scale \cite{Pierangeli2021_2}.
Our 3D-PELM has a record capacity $\mathcal C=362\times 362\times 120000\simeq 1.5 \times 10^{10}$,
more than one order of magnitude larger than any other optical processor.
Moreover, while increasing the capacity is challenging on many optical reservoir computers,
our 3D-PELM can be further scaled up by simply enlarging the measured volume of the optical field.

\begin{figure}[t!]
\centering
\hspace*{-0.8cm} 
\includegraphics[width=1.15\columnwidth]{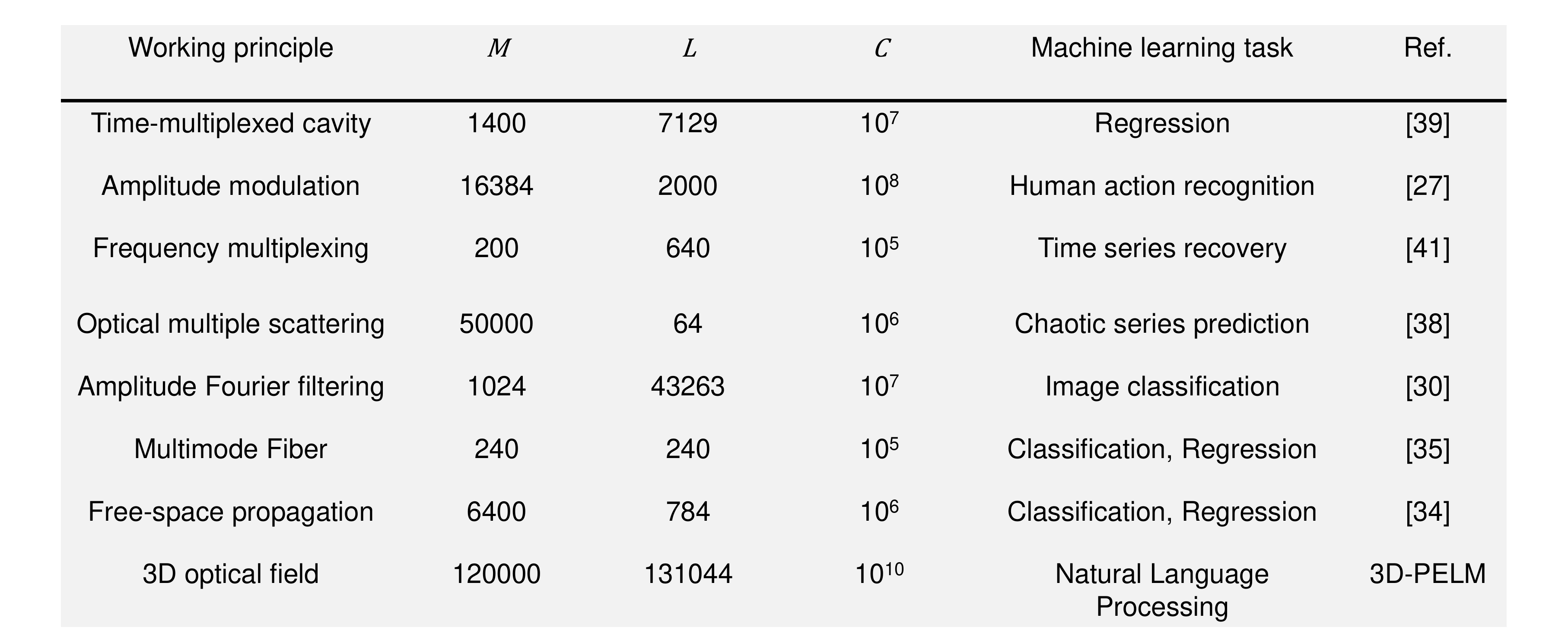}
\caption*{Table 1. Maximum network capacity of current photonic neuromorphic computing hardware.}
\label{Tab1}
\end{figure}

\begin{figure}[t!]
\centering
\hspace*{-0.1cm}
\includegraphics[width=0.7\columnwidth]{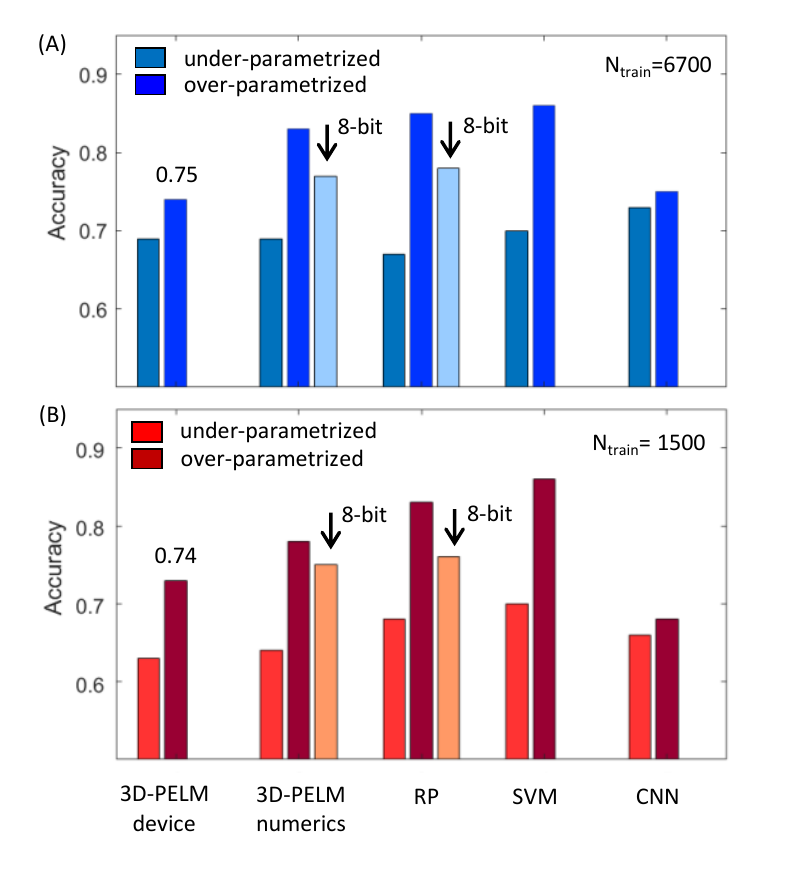}
\vspace*{-0.4cm}
\caption{Analysis of the IMDb accuracy.  (A-B) The comparison report the accuracy for the experimental device (3D-PELM device), 
the simulated device (3D-PELM numerics), the random projection method with ridge regression (RP), the support vector machine (SVM) 
and a convolutional nueral network (CNN) in both the under-parameterized ($M=1\times10^3$) and over-parametrized ($M=4\times10^4$ ) regime, 
for (A) $N_{\text{train}}=6700$,  and (B) $N_{\text{train}}=1500$. 8-bit numerical results, when applicable, refer to the over-parametrized regime.
}

\label{Figure5}
\end{figure}

\subsection*{\textbf{Comparison with digital machine learning}}

To further validate the capability of our photonic setup for text processing, 
we compare the accuracy of our 3D-PELM with various general-purpose digital neural networks on the IMDb task.
To underline the overall impact of the over-parameterization on the performance,
we consider two opposite parametrization regimes, respectively $M=1\times10^3$ and $M=4\times10^4$, 
and two distinct condition for the dataset size, $N_{\text{train}}=6700$ [Fig. 4(A)] and $N_{\text{train}}=1500$ [Fig. 4(B)].
Since the learning principle of the 3D-PELM is based on kernel methods (see Ref. \cite{Pierangeli2021}), 
we implement a support vector machine (SVM) and a ridge regression based on nonlinear random projetions (RP) \cite{Saade2016} 
as representative models of the kernel algorithms class.
The input for the SVM and the RP is the tfidf-encoded dataset, but no significant differences are found using the Hadamard dataset.
We also simulate the 3D-PELM, i.e., we evaluate our photonic scheme by using Eq. (2) to generate the output data.
A convolutional neural network (CNN) with the same number of trainable weights is used as an additional benchmark model.
Details on the various digital models are reported in Appendix E. 
The results, for a split ratio $N_{\text{train}}/N_{\text{test}} =0.67/0.33$, are shown in Figure 4.
Overall, we observe that the photonic device performs sentiment analysis on the IMDb with an accuracy comparable with standard digital methods.
The SVM sets the maximum expected accuracy (0.86).
The device over-parametrized accuracy (0.75 for $N_{\text{train}}=6700$)
is mainly limited by the limited precision of the optical components and noise,
and could be enhanced up to $0.83$ (3D-PELM numerics) by using a 64-bit camera.
In fact, when operating with 8-bit precision, both the simulated 3D-PELM and the RP model achieves performances
that agree well with the experiments.
Interestingly, Fig. 4(B) indicates that, for a limited set of training samples, 
the 3D-PELM surpasses the SVM and also the CNN. 
This points out an additional advantage that the photonic setup achieves thanks to over-parametrization.
Not only its accuracy is improved with respect the standard under-parametrized regime,
but the device can operates effectively in conditions where digital models are less accurate.

\section{DISCUSSION}

We have reported the first photonic implementation of natural language processing by performing sentiment analysis on the IMBd dataset. 
Our results demonstrate the feasibility of modern large-scale computing tasks in photonic hardware,
and can be potentially improved both in terms of performance,  
via low-noise high-resolution cameras and more complex encoding/training algorithms, and applications.
Further developments include the implementation of advanced tasks such as
multiclass sentiment analysis, which allows discriminating positive and negative sentences into different levels. 
Moreover, using different datasets such as the Stanford Sentiment Treebank (SST-5), we could train our 3D-PELM for providing ratings.
On the other hand, our versatile optical setting may allow implementing alternative strategies for photonic NLP, 
these including dimensionality reduction of the input data via optical random projection \cite{Miri2021}.
An optical scheme that preserves the sequential nature of language may be also conceived by introducing recurrent mechanisms in our 3D-PELM.
Another interesting direction is developing specific encoding schemes for representing text in the optical domain.
Viable approaches includes adapting bi-dimensional text representations such as Quick-Response (QR) codes. 
Furthermore, by following the concept of word embedding used in digital NLP \cite{Mikolov2013},
photonic hardware may be used to learn a purely optical word embedding, including semantics, word ordering, and syntax.

In conclusion, modern machine learning architectures rely on over-parametrization to reach the global optimum of objective functions. 
However, as better performances are associated with larger training sets, to reach an over-parametrized regime the number of adjustable parameters
has to grow accordingly, requiring in turn longer training times and more energy consumption \cite{Lacoste2019}.
This trend is evident in NLP: the state-of-the-art NLP model, GPT-3 \cite{Brown2020}, features 175 billions of parameters, in line with the exponential growth in the number of parameters of latest NLP models \cite{Narayanan2021}, and requires a dedicated supercomputing hardware for training. 
On the other hand, achieving over-parametrization for large-scale problems is challenging in photonic computing.
Here, we realize a novel photonic computing device that, thanks to its huge network capacity, naturally operates in the over-parametrized region.
The 3D-PELM is a very promising solution for developing a fully-scalable photonic learning platform.
Our photonic processor enables the first link between NLP and optical computing hardware, 
opening a new vibrant research stream towards fast and energetically efficient artificial intelligence tools.

\section*{APPENDIX A: PELM framework}
In ELMs a dataset with $N$ data points $\mathbf{X} = [(X_1,y_1),\dots, (X_N,y_N)]$, with $(X_i,y_i)\in \mathbb{R}^L\times \mathbb R$,
is mapped into a higher-dimensional feature space through a nonlinear transformation $g(\mathbf{X})$,
yielding the hidden-layer output matrix $\mathbf{H} = [g(X_1),\dots, g(X_N)]$, with $g(X_i)\in \mathbb R^M$.
A set of weights $\beta$ is learned via ridge regression such that the linear combination $\tilde{\mathbf{Y}}=\mathbf{H}\beta$
well approximates the true label vector $\mathbf{Y}=[y_1,\dots, y_N]$.
An explicit solution is $\beta = (\mathbf{H}^T\mathbf{H}+\lambda\,\mathbf{I})^{-1}\,\mathbf{H}^T\,\mathbf{Y}\,$
where $\lambda$ is the regularization parameter, $\mathbf{I}$ the identity. 
The use of ridge regression allows reducing at minimum the training costs,
which is crucial in applications requiring fast and reconfigurable learning.
In a free-space PELM, the nonlinear mapping of input data is realized by a combination of free-space optical propagation
and intensity detection \cite{Pierangeli2021}. 
The speckle-like pattern detected by a camera forms the nonlinear features of the input sample. 
The optical mapping reads as $\mathbf{H} = G[\mathbf{M}\,\exp{i(\mathbf{X}+\mathbf{W})}]$, 
where the input dataset $\mathbf{X}$ is phase-encoded via spatial light modulation. $\mathbf{W}$ is the embedding matrix, 
$G$ the detector response function, and $\mathbf M$ the discrete Fourier transform that models propagation to the focal plane.
For more details on the PELM architecture and its working principle see Ref. \cite{Pierangeli2021}.

\section*{APPENDIX B: Experimental setup}
The optical setup of the 3D-PELM is sketched in Figure~1(B). A CW laser beam (532 nm, $100$ mW) 
is expanded and collimated onto a phase-only SLM (Hamamatsu X13138, 1280 $\times$ 1024 pixels, 60 Hz, 8 bits),
that encodes data on the optical wavefront.
The SLM operates in the $[0, 2 \pi]$ interval with a linear response function.
Both the input and the embedding signal are encoded within the $[0, \pi]$ range and superimposed.
After free-space propagation through a lens ($f=150$mm),
the far-field three-dimensional intensity distribution is accessed simultaneously 
by three separated imaging systems and detected by three cameras (Basler Ace 2 Pro, 1920 $\times$ 1200 pixels, 12 bits).
The speckle-like field spatially decorrelates during propagation; 
the transverse intensity is transformed in a random way from one plane to the other, 
provided that these planes are at distances larger than the longitudinal field correlation length.
To carry out the measurements, the propagating beam is divided by a series of beam splitters into separated imaging paths
with different tunable optical paths. 
The far-field planes are selected in order to sample three speckle-like distribution having a negligible mutual intensity correlation.
This is confirmed by intensity correlation measurements. 
Every additional plane contains additional information. In fact, keeping fixed $M$, we find an improvement in the classification accuracy
when using three far-field planes instead of a single intensity distribution.
In this condition, each camera maps a distinct portion of the high-dimensional feature space in which the input field is projected.

The measured set of speckle-like patterns is elaborated on a conventional computer.
Speckle grains typically extend over adjacent pixels [Fig. 1(C)]. 
To avoid spatial correlations within the single speckle pattern, 
pixels are grouped into square modes (channels) and the average intensity on each channels forms the output signals.

\section*{APPENDIX C: 3D-PELM training}
Training operates by loading the randomly-ordered input dataset on the SLM and measuring three speckle-like intensity distributions 
for each input sample (paragraph). From the acquired signals we randomly select $M$ of the available channels, 
and we use the corresponding $M$ intensity values to form the output dataset.
This dataset is split randomly into a training and test dataset by using a split ratio $N_{\text{train}}/N_{\text{test}} =0.67/0.33$, 
which is kept fixed throughout the analysis. All the classification accuracies we obtained refer to this hyperparameter.
The training dataset is the output matrix $\mathbf{H}$ used for training (see Appendix A). 
Since our task corresponds to a binary classification problem, the target labels are $y_i \in \{0,1\}$.
According to the PELM framework, training by ridge regression consists in solving the equation for the $\beta$ vector. 
This matrix inversion is performed using the ML Python Library {\it Sci-kit learn\/}. 
The regularization parameter $\lambda$ is varied in the $[10^{-4}, 10^2]$ range. 
As its value does not influence considerably the testing accuracy, we use $\lambda=10^{-4}$.  
Given a predicted output $\hat y_i$, 
the predicted label is given by $\tilde{y}_i= \Theta(\hat y_i - 0.5)$, where $\Theta$ is the Heaviside step function. 
Text classification performances are evaluated via the accuracy $\sum_i \left[ y_i=\tilde{y}_i \right] /N$,
where $[\cdot]$ are the Iverson bracket defined as $[P]=1$ if $P$ is true and $0$ otherwise.

\section*{APPENDIX D: NLP task, text pre-processing and optical encoding}
We consider the IMDb dataset \cite{Maas2011}, a widely used benchmark for polarity-based text classification \cite{Rehamn2019}.
The dataset consists of $5\times10^4$ movie reviews, each one forming the $i$-th input data sample. 
Each review is made of a paragraph consisting of multiple sentences, and expresses an overall positive or negative sentiment (target variable). 
This sentiment analysis task hence consists in a large-scale binary classification.
To implement the problem, the whole set of IMDb sentences is pre-processed by removing special characters (e.g. html), 
lowercasing, removing stop words and lemmatizing. 
The length of the resulting vector can be varied by neglecting words that appears in the corpus with a frequency lower than a given threshold. 
To compute the Walsh-Hadamard transform and obtain a dense text representation,
zero-padding is used to reach a vector length $d = \left \lceil{\log_2 V }\right \rceil $.
For example, results in Fig. 2 refer to data with $V = 10^4$ that are padded to $L=2^{14}= 16384$ 
and reshaped as a $128 \times 128$ square matrix, 
while the full vocabulary ($V= 131044$) requires zero-padding and paragraphs are arranged on $362 \times 362$ input modes (Fig. 3). 
After WHT computation, the values are normalized in the $[0, \pi]$ range and loaded on the SLM.
Each element of the input $\mathbf{X}_{\text{WHT}}$ is displayed onto a square block of SLM pixels.
The embedding matrix $\mathbf{W}$ is a uniformly distributed random matrix for all the experiments. 
Its values $w_{ij}$ are selected in $[0, \pi]$ before the training and kept fixed for a given photonic computation.

\section*{APPENDIX E: Digital machine learning}
The 3D-PELM is numerically simulated by following the optical mapping model in Eq. (1) (see also Ref. \cite{Pierangeli2021}).
Data encoding and training operates in the same way as the photonic device, and the output matrix is computed by using Eq. (2)
with $G(z)=\vert z\vert^2$ (quadratic nonlinearity).
In the RP model, data projection in the feature space is obtained by matrix multiplication with a real-valued uniformly-distributed 
random matrix and using a quadratic activation function. Training occurs by ridge regression.
For simulating the 8-bit machines, the input data and the output values forming the linear readout layer 
are properly discretized to values with 8-bit precision. 
The SVM is a binary classifier and thus it is especially suited for the IMDb task.
We use the {\it Sci-kit learn\/} built-in function on the tfid-processed dataset and 
the over-parametrized regime is evaluated by using $M=10^3$.
The CNN is composed by four convolutional layer with nonlinear activation and a fully connected layer.
The CNN scheme is designed to have $M$ total nodes, and its accuracy does not depend significatively on the network details and hyperparameters.

\vspace*{0.2cm}

\textbf{Funding.} We acknowledge funding from the Italian Ministry of Education, University and Research (PRIN PELM 20177PSCKT),
Sapienza Research, and Enrico Fermi Research Center (CREF).

\textbf{Author Contributions.} D.P. and C.C. conceived the research. C.M.V. proposed application to NLP. 
D.P. developed the 3D photonic network.  I.G. and D.P. carried out experimental measurements. 
C.M.V. and I.G. performed data analysis.  D.P and C.C. co-supervised the project. All authors contributed to the manuscript.

\textbf{Acknowledgments.} We thank I. MD Deen, V.H. Santos, and F. Farrelly for technical support in the laboratory.

\textbf{Disclosures.} The authors declare no conflicts of interest.

\textbf{Data availability.} The data that support the results presented in this study are available from the corresponding author upon reasonable request.

\textbf{Code availability.} The codes developed within this research are available from the corresponding author upon reasonable request.


\vspace*{-0.1cm}
\begin{thebibliography}{70}


\bibitem{Narayanan2021} D. Narayanan et al., Efficient large-scale language model training on GPU clusters using megatron-LM, arXiv:2104.04473 (2021).
\bibitem{Strubell2020} E. Strubell, A. Ganesh, and A. McCallum, Energy and policy considerations for modern deep learning research,
{\it Proc. AAAI Conf. on Artif. Intell. \/}{\bf 34}, 13693–13696 (2020).
\bibitem{Chatelain2021} A. Chatelain, A. Djeghri, D. Hesslow, J. Launay, and I. Poli, Is the number of trainable parameters all that actually
matters?, arXiv:2109.11928 (2021).
\bibitem{Thompson2020} N.C. Thompson, K. Greenewald, K. Lee, and G.F. Manso, The computational limits of deep learning,	arXiv:2007.05558 (2020).
\bibitem{Wetzstein2020} G. Wetzstein, A. Ozcan, S. Gigan, S. Fan, D. Englund, M. Soljačić, C. Denz, D.A.B. Miller and D. Psaltis, Inference in artificial intelligence with deep optics and photonics, {\it Nature\/} {\bf 588}, 39-47 (2020).
\bibitem{Shastri2021} B.J. Shastri, A.N. Tait, T.F. de Lima, W.H.P. Pernice, H. Bhaskaran, C.D. Wright, and P.R. Prucnal, Photonics
for artificial intelligence and neuromorphic computing, {\it Nat. Photonics \/}{\bf 15}, 102–114 (2021).
\bibitem{Zhou2022} H. Zhou et al., Photonic matrix multiplication lights up photonic accelerator and beyond,
{\it Light Sci. Appl.\/}{ \bf 11}, 1-21 (2022).

\bibitem{Xu2021} X. Xu, M. Tan, B. Corcoran, J. Wu, A. Boes, T.G. Nguyen, S.T. Chu, B.E. Little, D.G. Hicks, R. Morandotti, A. Mitchell, and D.J. Moss, 11 TOPS photonic convolutional accelerator for optical neural networks, {\it Nature\/} {\bf 589}, 44-51 (2021).
\bibitem{McMahon2021} T. Wang, S.-Y. Ma, L.G. Wright, T. Onodera, B. Richard, and P.L. McMahon, An optical neural network using less
than 1 photon per multiplication, {\it Nat. Commun. \/}{\bf 13}, 1-8 (2022).
\bibitem{Shen2017} Y. Shen, N.C. Harris, S. Skirlo, M. Prabhu, T. Baehr-Jones, M. Hochberg, X. Sun, S. Zhao, H. Larochelle, D. Englund, and M. Soljacic, 
Deep learning with coherent nanophotonic circuits, {\it Nat. Photon. \/} {\bf 11}, 441–446 (2017).
\bibitem{Tait2017} A.N. Tait , T.F. de Lima, E. Zhou, A.X. Wu, M-A. Nahmias, B.J. Shastri, P.R. Prucnal, 
Neuromorphic photonic networks using silicon photonic weight banks, {\it Sci. Rep. \/} {\bf 7}, 7430 (2017).
\bibitem{Feldmann2019} J. Feldmann, N. Youngblood, C.D. Wright, H. Bhaskaran and W.H.P. Pernice, All-optical spiking neurosynaptic networks with self-learning capabilities, {\it Nature\/} {\bf 569}, 208-214 (2019).
\bibitem{Stelzer2021} F. Stelzer, A. Röhm, R. Vicente, I. Fischer, and S. Yanchuk, Deep neural networks using a single neuron:
folded-in-time architecture using feedback-modulated delay loops,{\it Nat. Commun. \/}{\bf 12} (2021).
\bibitem{Feldmann2021} J. Feldmann,  N. Youngblood, M. Karpov, H. Gehring, X. Li, M. Stappers, M. Le Gallo, X. Fu, A. Lukashchuk, A.S. Raja, J. Liu, C.D. Wright, A. Sebastian, T.J. Kippenberg, W.H.P. Pernice, and H. Bhaskaran, Parallel convolutional processing using an integrated photonic tensor core,
{\it Nature \/} {\bf 589}, 52-58 (2021).
\bibitem{Engheta2019} N. Mohammadi Estakhri, B. Edwards, and N. Engheta, Inverse-designed metastructures that solve equations, 
{\it Science\/} {\bf 363}, 1333 (2019).
\bibitem{Lin2018} X. Lin, Y. Rivenson, N. T. Yardimci, M. Veli, Y. Luo, M. Jarrahi, and A. Ozcan, All-optical machine learning using diffractive deep neural networks, {\it Science \/}{\bf 361}, 1004–1008 (2018).
\bibitem{Zhou2021} T. Zhou, X. Lin, J. Wu, Y. Chen, H. Xie, Y. Li, J. Fan, H. Wu, L. Fang, and Q. Dai, Large-scale neuromorphic
optoelectronic computing with a reconfigurable diffractive processing unit, {\it Nat. Photonics\/} {\bf 15}, 367–373 (2021).


\bibitem{Verstraeten2007} D. Verstraeten, B. Schrauwen, M. D’Haene, and D. Stroobandt, An experimental unification of reservoir computing
methods, {\it Neural Networks \/}{\bf 20}, 391–403 (2007).
\bibitem{Huang2012} G.B. Huang, H. Zhou, X. Ding, and R.Zhang, Extreme Learning Machine for Regression and Multiclass Classification, {\it IEEE 
Transactions on Systems, Man, and Cybernetics, Part B (Cybernetics) \/}{\bf 42}, 513-529 (2012).

\bibitem{VanDerSande2017} G. Van der Sande, D. Brunner, and M.C. Soriano, Advances in photonic reservoir computing, {\it Nanophotonics \/} {\bf 6}, 561–576 (2017).
\bibitem{Brunner2013} D. Brunner, M.C. Soriano, C. Mirasso, I. Fischer, Parallel photonic information processing at gigabyte per second data rates using transient states, {\it Nat. Commun. \/} {\bf 4}, 1364~(2013).
\bibitem{Vinkier2015} Q. Vinckier, F. Duport, A. Smerieri, K. Vandoorne, P. Bienstman, M. Haelterman, and S. Massar, High-Performance Photonic Reservoir Computer Based on a Coherently Driven Passive Cavity, {\it Optica \/} {\bf 2}, 438 (2015).
\bibitem{Larger2017} L. Larger, A. Baylón-Fuentes, R. Martinenghi, V.S. Udaltsov, Y.K. Chembo, and M. Jacquot, High-speed photonic reservoir computing using a time-delay based architecture: million words per second classification, {\it Phys. Rev. X \/} {\bf 7}, 011015 (2017).
\bibitem{Ballarini2020} D. Ballarini et al., Polaritonic neuromorphic computing outperforms linear classifiers, {\it Nano Lett.\/}{\bf 20}, 3506–3512 (2020).
\bibitem{Saade2016} A. Saade, F. Caltagirone, I. Carron , L. Daudet, A. Dremeau, S. Gigan, F. Krzakala, 
Random projections through multiple optical scattering: approximating kernels at the speed of light, 
{\it IEEE International Conference on Acoustics, 
Speech and Signal Processing (ICASSP)\/} 6215–6219 (2016).
\bibitem{Bueno2018} J. Bueno, S. Maktoobi, L. Froehly, I. Fischer, M. Jacquot, L. Larger, and D. Brunner, Reinforcement learning in a large-scale photonic recurrent neural network, {\it Optica\/} {\bf 5}, 756–760 (2018).
\bibitem{Antonik2019} P. Antonik, N. Marsal, D. Brunner, and D. Rontani, Human action recognition with a large-scale brain-inspired photonic computer,
{\it Nat. Mach. Intell. \/}{\bf 1}, 530–537~(2019).
\bibitem{Rohm2020} A. Röhm, L. Jaurigue, and K. L\"udge, Reservoir Computing Using Laser Networks, {\it IEEE J. Sel. Top. Quantum Electron.\/}
{\bf 26}, 1 (2020).
\bibitem{Paudel2020} U. Paudel, M. Luengo-Kovac, J. Pilawa, T.J. Shaw, and G.C. Valley, Classification of time-domain waveforms using a speckle-based optical reservoir computer, {\it Opt. Express\/} {\bf 28}, 1225 (2020).
\bibitem{Miscuglio2020} M. Miscuglio, Z. Hu, S. Li, J. George, R. Capanna, P.M. Bardet, P. Gupta, and V.J. Sorger, Massively Parallel Amplitude-Only Fourier Neural Network, {\it Optica\/} {\bf 7}, 1812 (2020).
\bibitem{Sunada2021} S. Sunada and A. Uchida, Photonic neural field on a silicon chip: large-scale, high-speed neuro-inspired computing
and sensing, {\it Optica\/} {\bf 8}, 1388 (2021).
\bibitem{Pavesi2021} M. Borghi, S. Biasi, and L. Pavesi, Reservoir computing based on a silicon microring and time multiplexing for
binary and analog operations, {\it Sci. Rep.\/} {\bf 11}, 15642 (2021).
\bibitem{Porte2021} X. Porte, A. Skalli, N. Haghighi, S. Reitzenstein, J. A. Lott, and D. Brunner, A complete, parallel and autonomous
photonic neural network in a semiconductor multimode laser, {\it J. Physics: Photonics \/}{\bf 3}, 024017 (2021).


\bibitem{Pierangeli2021} D. Pierangeli, G. Marcucci, and C. Conti, Photonic extreme learning machine by free-space optical propagation,
{\it Photonics Res. \/} {\bf 9}, 1446 (2021).
\bibitem{Psaltis2021} U. Teğin, M. Yıldırım, İ. Oğuz, C. Moser, and D. Psaltis, Scalable optical learning operator,
 {\it Nat. Comput. Sci.\/} {\bf 1}, 542–549 (2021).
\bibitem{Pierangeli2021_2} D. Pierangeli, G. Marcucci, and C. Conti, (2021, August). Neuromorphic computing device using optical shock waves,
In {\it OSA Nonlinear Optics\/} 2021 (pp. NTh1A-3), OSA Technical Digest (Optica Publishing Group, 2021).
\bibitem{Ciuti2022} Z. Denis, I. Favero , and C. Ciuti, Photonic Kernel Machine Learning for Ultrafast Spectral Analysis,
 {\it Phys. Rev. Appl.\/}{\bf 17}, 034077 (2022).

\bibitem{Rafayelyan2020} M. Rafayelyan, J. Dong, Y. Tan, F. Krzakala, and S. Gigan, Large-Scale Optical Reservoir Computing for Spatiotemporal Chaotic Systems Prediction, {\it Phys. Rev. X \/}{ \bf 10}, 041037 (2020).
\bibitem{Ortin2015} S. Ortín, M. C. Soriano, L. Pesquera, D. Brunner, D. San-Martín, I. Fischer, C. R. Mirasso, and J. M. Gutiérrez, 
A unified framework for reservoir computing and extreme learning machines based on a single time-delayed neuron, {\it Sci. Reports\/} {\bf 5} (2015).

\bibitem{Matuszewski2021} R. Mirek, A. Opala, P. Comaron, M. Furman, M. Król, K. Tyszka, B. Seredyński, D. Ballarini, D. Sanvitto, T.C.H.
Liew, W. Pacuski, J. Suffczyński, J. Szczytko, M. Matuszewski, and B. Piętka, Neuromorphic binarized polariton networks, 
{\it Nano Lett.\/} {\bf 21}, 3715–3720 (2021).
\bibitem{Lupo2021} A. Lupo and S. Massar, Parallel extreme learning machines based on frequency multiplexing, {\it Appl. Sci.\/} {\bf 12}, 214 (2021).

\bibitem{Marcucci2020} G. Marcucci, D. Pierangeli, and C. Conti, Theory of neuromorphic computing by waves: machine learning by rogue waves, dispersive shocks, and solitons,  {\it Phys. Rev. Lett.\/} {\bf 125}, 093901 (2020).
\bibitem{Zhang2021} H. Zhang at al., An optical neural chip for implementing complex-valued neural network, {\it Nat. Commun.\/} {\bf 12}, 457~(2021).

\bibitem{Maas2011} A. L. Maas, R. E. Daly, P. T. Pham, D. Huang, A. Y. Ng, and C. Potts, Learning word vectors for sentiment analysis,
in {\it Proceedings of the 49th Annual Meeting of the Association for Computational Linguistics: Human Language
Technologies\/}, 142–150 (Association for Computational Linguistics, Portland, Oregon, USA, 2011).
\bibitem{Belkin2019} M. Belkin, D. Hsu, S. Ma, and S. Mandal, Reconciling modern machine-learning practice and the classical
bias–variance trade-off, {\it Proc. Natl. Acad. Sci. \/}{\bf 116}, 15849–15854 (2019).
\bibitem{Advani2020} M. S. Advani, A. M. Saxe, and H. Sompolinsky, High-dimensional dynamics of generalization error in neural
networks,{\it Neural Networks \it} {\bf 132}, 428–446 (2020).
\bibitem{Montanari2020} S. Mei and A. Montanari, The generalization error of random features regression: Precise asymptotics and double
descent curve, arXiv:1908.05355 (2020).

\bibitem{Babic2020} K. Babić, S. Martinčić-Ipšić, and A. Meštrović, Survey of neural text representation models, {\it Information} {\bf 11}, 511 (2020).
\bibitem{Ashrafi2017} A. Ashrafi, Walsh–hadamard transforms: A review, in {\it Advances in Imaging and Electron Physics\/}, 1-55 (Elsevier, 2017).
\bibitem{Soltanolkotabi2019} M. Soltanolkotabi, A. Javanmard, and J. D. Lee, Theoretical insights into the optimization landscape of overparameterized
shallow neural networks, {\it IEEE Transactions on Inf. Theory\/} {\bf 65}, 742–769 (2019).

\bibitem{Pierangeli2021_2} D. Pierangeli, M. Rafayelyan, C. Conti, and S. Gigan, Scalable Spin-Glass Optical Simulator,
 {\it Phys. Rev. Appl.} {\bf 15}, 034087 (2021).
\bibitem{Miri2021} M.-A. Miri, Integrated random projection and dimensionality reduction by propagating light in photonic lattices,
 {\it Opt. Lett \/} {\bf 46}, 4936 (2021).

\bibitem{Mikolov2013} T. Mikolov, K. Chen, G. Corrado, and J. Dean, Efficient estimation of word representations in vector space, arXiv:1301.3781 (2013).
\bibitem{Lacoste2019} A. Lacoste, A. Luccioni, V. Schmidt, and T. Dandres, Quantifying the carbon emissions of machine learning, arXiv:1910.09700 (2019).
\bibitem{Brown2020} T.B. Brown et al.,	Language Models are Few-Shot Learners, arXiv:2005.14165 (2020).

\bibitem{Rehamn2019} A.U. Rehman et al., A hybrid CNN-LSTM model for improving accuracy of movie reviews sentiment analysis, {\it Multimedia Tools and Applications\/} {\bf 78}, 26597 (2019).





\end{thebibliography}
\end{document}